\documentstyle[prl,aps]{revtex}
\begin{document}
\draft

\title{Exact Momentum Distribution of a Fermi Gas in One Dimension}
\author{Girish S. Setlur}
\address{Department of Physics and Materials Research Laboratory
\\ University of Illinois at Urbana-Champaign, Urbana, Illinois 61801}

\date{\today}
\maketitle


\begin{abstract}
                        
 We introduce an exactly solvable model of a fermi gas in one dimension
 and compute the momentum distribution exactly. This is based on a
 generalisation of the ideas of bosonization in one dimension.
 It is shown that in the RPA limit(the ultra-high density limit)
 the answers we get are the exact answers for a homogeneous fermi gas
 interacting via a two-body repulsive coulomb interaction.
 Furthermore, the solution may be obtained exactly for arbitrary functional
 forms of the interaction, so long as it is purely repulsive. No linearization
 of the bare fermion dispersion is required. We find that for the
 interaction considered, the fermi surface is intact for weak repulsion and
 is destroyed only for sufficiently strong repulsion. Comparison with other
 models like the supersymmetric t-J model with inverse square
 interactions is made.

\end{abstract}

 Recent years have seen remarkable developments in many-body theory
 in the form an assortment of techniques that may be loosely termed
 bosonization. The beginnings of these types of techiniques may be
 traced back to the work of Tomonaga\cite{Tom}
 and later on by Luttinger \cite{Lutt} and by Lieb and Mattis
 \cite{Lieb}. The work of Sawada \cite{Sawada} and Arponen and Pajanne
 \cite{Arponen} in recasting the fermi gas problem in a bose language has
 to be mentioned.  Arponen and Pajanne recover corrections to 
 the Random Phase Approximation (RPA) of Bohm and Pines
 \cite{Bohm} in a systematic manner. 
 In the 70's an attempt was made by Luther \cite{Luther}
 at generalising these ideas to higher dimensions. Closely related to this
 is work by Sharp et. al. \cite{Sharp} in current algebra.
 More progress was made by Haldane \cite{Haldane} which culminated in
 the explicit computation of the single particle propagator by
 Castro-Neto and Fradkin \cite{Neto} and by Houghton, Marston et.al.
 \cite{Mars} and also by Kopietz et. al. \cite{Kop}.
 Rigorous work by Frohlich and Marchetti\cite{Froch}
 is also along similar lines. Also the work of Frau et. al. \cite{Zemba}
 on algebraic bosonization is relevant to the present Letter as the
 authors have considered effects beyond the linear dispersion in
 that article.

 In this Letter, we try to use the
 formalism introduced by us \cite{Setlur1} in an earlier preprint
  to write down an interaction \cite{Setlur2},\cite{Setlur3} that
 closely mimics the two-body repulsive interaction between the fermions
 and compute the momentum distribution exactly. It is also shown that in
 the ultra-high density limit, the interaction is exactly the two-body
 repulsive coulomb interaction insofar as it is able to recover the random
 phase approximation exactly.
 This momentum distribution
 differs from the traditional Luttinger model in one dimension
 in this aspect, namely,
 the discontinuity at the fermi momentum is present for weak
 repulsion and is destroyed only for stronger repulsions. But the present
 model does share similarities with other models
 such as the supersymmetric t-J model \cite{Kura} which does show a 
 discontinuity at the fermi moementum in its momentum distribution. 

 In the sissa preprint \cite{Setlur1} we wrote down a correspondence
 between products
 of fermi fields and bose fields that represent the displacements of the
 fermi sea, in a manner analogous to the fermi-surface displacements introduced
 by Haldane \cite{Haldane}. There, the bose fields in question were written
 as $ a_{ {\bf{k}} }({\bf{q}}) $ and $ a^{\dagger}_{ {\bf{k}} }({\bf{q}}) $ 
 Linear combination of these operators would then correspond to displacement
 oprators of the fermi sea at wavevector $ {\bf{k}} $ by an amount $ {\bf{q}} $.
 This is a generalisation of the concept of fermi-surface displacements where
 the magnitude of $ {\bf{k}} $ would be restricted to equal $ k_{f} $ the fermi
 momentum. This enables us to draw up a correspondence as described in the
 sissa preprint  \cite{Setlur1}. The boldface on the vectors $ {\bf{q}} $ and
 $ {\bf{k}} $ serve to illustrate that these ideas admit a straightforward
 generalisation to more than one dimension.
Let $ c_{ {\bf{k}} } $ and
 $ c^{\dagger}_{ {\bf{k}} } $ be the fermi fields in question, then,
\[
          c^{\dagger}_{ {\bf{k+q/2}} }c_{ {\bf{k-q/2}} }
 = n_{F}({\bf{k}})\frac{N}{\langle N \rangle }\delta_{ {\bf{q}}, {\bf{0}} }
 + (\sqrt{ \frac{N}{\langle N \rangle } })
[\Lambda_{ {\bf{k}} }({\bf{q}})a_{ {\bf{k}} }(-{\bf{q}})
 + \Lambda_{ {\bf{k}} }(-{\bf{q}})a^{\dagger}_{ {\bf{k}} }({\bf{q}}) ]
\]
\[
 +
 \sum_{ {\bf{q}}_{1} }\Lambda_{ {\bf{k+q/2 - q_{1}/2}} }(-{\bf{q_{1}}})
\Lambda_{ {\bf{k - q_{1}/2}} }({\bf{q-q_{1}}})
  a^{\dagger}_{ {\bf{k}} + {\bf{q}}/2 - {\bf{q}}_{1}/2 }({\bf{q}}_{1})
a_{ {\bf{k}} -  {\bf{q}}_{1}/2 }(-{\bf{q}} + {\bf{q}}_{1})
\]
\begin{equation}
 - \sum_{ {\bf{q}}_{1} }\Lambda_{ {\bf{k-q/2 + q_{1}/2}} }(-{\bf{q_{1}}})
\Lambda_{ {\bf{k + q_{1}/2}} }({\bf{q-q_{1}}})
  a^{\dagger}_{ {\bf{k}} - {\bf{q}}/2 + {\bf{q}}_{1}/2 }({\bf{q}}_{1})
a_{ {\bf{k}} + {\bf{q}}_{1}/2 }(-{\bf{q}} + {\bf{q}}_{1})
\label{FORM}
\end{equation}
Here, $ a_{ {\bf{k}} }({\bf{q}}) $ and $ a^{\dagger}_{ {\bf{k}} }({\bf{q}}) $
 are exact bose operators and $ c_{ {\bf{k}} } $ and
 $ c^{\dagger}_{ {\bf{k}} } $ are exact fermion operators. In symbols,
\begin{equation}
\{c_{ {\bf{k}} }, c_{ {\bf{k}}^{'} } \} = 0; \mbox{   }
\{c_{ {\bf{k}} }, c^{\dagger}_{ {\bf{k}}^{'} } \} 
= \delta_{ {\bf{k}}, {\bf{k}}^{'} }
\end{equation}
and,
\begin{equation}
[a_{ {\bf{k}} }({\bf{q}}), a_{ {\bf{k}}^{'} }({\bf{q}}^{'}) ] = 0; \mbox{   }
[a_{ {\bf{k}} }({\bf{q}}), a^{\dagger}_{ {\bf{k}}^{'} }({\bf{q}}^{'}) ]
= \delta_{ {\bf{k}}, {\bf{k}}^{'} }\delta_{ {\bf{q}}, {\bf{q}}^{'} }
\end{equation}
and, 
$ \Lambda_{ {\bf{k}} }({\bf{q}}) = \sqrt{ n_{F}({\bf{k+q/2}})(1 - n_{F}({\bf{k-q/2}})) } $
and, $ n_{F}({\bf{k}}) = \theta(k_{f}-k) $.
 Furthermore, the kinetic energy operator is given in the fermi language as,
\begin{equation}
 K = \sum_{ {\bf{k}} }\epsilon_{ {\bf{k}} }c^{\dagger}_{ {\bf{k}} }
c_{ {\bf{k}} }
\end{equation}
where $ \epsilon_{ {\bf{k}} } = k^{2}/2m $. The same operator
has an extremely elegant form in the bose language,
\begin{equation}
K = E_{0} + \sum_{ {\bf{k}}, {\bf{q}} }\omega_{ {\bf{k}} }({\bf{q}})
a^{\dagger}_{ {\bf{k}} }({\bf{q}})
a_{ {\bf{k}} }({\bf{q}})
\end{equation}
here, $ E_{0} = \sum_{ {\bf{k}} }\epsilon_{ {\bf{k}} }n_{F}({\bf{k}}) $,
 and the excitation energy is given by,
\begin{equation}
\omega_{ {\bf{k}} }({\bf{q}}) = (\frac{ {\bf{k.q}} }{m})
\Lambda_{ {\bf{k}} }(-{\bf{q}})
\end{equation}
 Also the number operator 
$ N = \sum_{ {\bf{k}} }c^{\dagger}_{ {\bf{k}} }c_{ {\bf{k}} } $ is to be
 distinguished from its expectation value, 
$ \langle N \rangle = \sum_{ {\bf{k}} }n_{F}({\bf{k}}) $.
 The coefficient $ \Lambda_{ {\bf{k}} }(-{\bf{q}}) $ is non-zero only when
 $ {\bf{k.q}} \geq 0 $.
 The meaning of the above formula in Eq.(~\ref{FORM})
  is as follows. All the dynamical moments
 of $ c^{\dagger}_{ {\bf{k+q/2}} }c_{ {\bf{k+q/2}} } $ evaluated in the
 fermi language are going to be identical to those evaluated using the
 bose representation of this product, provided, we identify the noninteracting
 fermi-sea with the bose vacuum. In other words,
\begin{equation}
a_{ {\bf{k}} }({\bf{q}})|FS\rangle = 0
\end{equation} 
 From this it is possible to recast any problem involving interacting fermions
 in a form involving only bose fields. In particular, if we selectively choose
 parts of the interaction terms written out in the bose language, it may be
 possible to compute exactly, important quantities such as, momentum
 distribution, dynamical density correlation functions and so on.
 Let us now focus on the homogeneous fermi gas. The interaction has the
 form,
\begin{equation}
U = \sum_{ {\bf{q}} \neq 0 }
\frac{ v_{ {\bf{q}} } }{2V}
(\rho_{ {\bf{q}} }\rho_{ -{\bf{q}} } - N)
\end{equation}
 Written out in terms of the bose fields, it has a part that is
 quadratic in the bose fields, and we have cubic and quartic terms as well.
 Let us just focus
 on the part that is quadratic in the bose fields. This amounts to postulating
 a phenomenological interaction that is not exactly of the the two-body
 type (in the original fermi system) but something that mimics it.
 How closely does it mimic this two-body interaction, is the 
 big question.
 This question may be answered by computing the dielectric function and 
 demonstrating that it is exactly equal to the RPA dielctric function. 
 This has been done in our important preprint that computes the
 single-particle green functions of the same system\cite{Setlur3}.
 Thus our results are
 the exact results for a homogeneous fermi gas with two-body repulsive
 interactions in the same limit in which the RPA is exact.
 Written out in the bose language this interaction has the form,
\begin{equation}
H_{I} = \sum_{ {\bf{q}} \neq 0 }\frac{ v_{ {\bf{q}} } }{2V}
\sum_{ {\bf{k}}, {\bf{k}}^{'} }
[\Lambda_{ {\bf{k}} }({\bf{q}})a_{ {\bf{k}} }(-{\bf{q}}) + 
 \Lambda_{ {\bf{k}} }(-{\bf{q}})a^{\dagger}_{ {\bf{k}} }({\bf{q}})]
[\Lambda_{ {\bf{k}}^{'} }(-{\bf{q}})a_{ {\bf{k}}^{'} }({\bf{q}}) +
 \Lambda_{ {\bf{k}}^{'} }({\bf{q}})a^{\dagger}_{ {\bf{k}}^{'} }(-{\bf{q}})]
\end{equation}
 The full hamiltonian is now $ H = K + H_{I} $. This hamiltonian
 may be diagonalised exactly and the momentum distribution may be computed.
 Alternatively, the equation of motion method may be used to obtain the
 same answers.
 In fact, we have used both these methods \cite{Setlur3}
 to obtain the exact answer for the momentum distribution in one
 dimension (the formulas written down till now are valid in any number
 of dimensions). The exact momentum distribution found by us has the form
 ($ \hbar = c = 1 $),
\[
\langle c^{\dagger}_{ k }c_{ k } \rangle
 = n_{F}(k) +
(2\pi k_{f})
\int_{-\infty}^{+\infty} \mbox{ }\frac{ dq_{1} }{2\pi}\mbox{ }
\frac{ \Lambda_{ k - q_{1}/2 }(-q_{1}) }
{ 2\omega_{R}(q_{1})(\omega_{R}(q_{1}) + \omega_{k - q_{1}/2}(q_{1}))^{2}
(\frac{ m^{3} }{q_{1}^{4}})( cosh(\lambda(q_{1})) - 1 ) }
\]
\begin{equation}
- (2\pi k_{f})
\int_{-\infty}^{+\infty} \mbox{ }\frac{ dq_{1} }{2\pi}\mbox{ }
\frac{ \Lambda_{ k + q_{1}/2 }(-q_{1}) }
{ 2\omega_{R}(q_{1})(\omega_{R}(q_{1}) + \omega_{k + q_{1}/2}(q_{1}))^{2}
(\frac{ m^{3} }{q_{1}^{4}})( cosh(\lambda(q_{1})) - 1 ) }
\label{SOLN}
\end{equation}
\begin{equation}
\lambda(q) = (\frac{2 \pi q}{m})(\frac{1}{v_{q}})
\end{equation}
\begin{equation}
\omega_{R}(q) = (\frac{ |q| }{m})
\sqrt{ \frac{ (k_{f} + q/2)^{2}  - (k_{f} - q/2)^{2}exp(-\lambda(q)) }
{ 1 - exp(-\lambda(q)) } }
\end{equation}
 The above answer for the momentum distribution Eq.(~\ref{SOLN}) is the
 exact momentum distribution of a homogeneous fermi gas in one dimension
 interacting via a two-body repulsive interaction.
 As has been pointed out earlier, the RPA dielctric function as descibed by
 Bohm and Pines\cite{Bohm} may be recovered exactly by this model
 provided one identifies the density operator with the object 
 obtained by retaing only terms linear in the sea-displacements.
\begin{equation}
\rho_{ {\bf{q}} } = \sum_{ {\bf{k}} }[\Lambda_{ {\bf{k}} }({\bf{q}})
a_{ {\bf{k}} }(-{\bf{q}}) + \Lambda_{ {\bf{k}} }(-{\bf{q}})
a^{\dagger}_{ {\bf{k}} }({\bf{q}})]
\end{equation}
 The above form ignores the terms quadratic in the sea-displacements. The
 justification being that this definition along the with the standard
 procedure for obtaining the dielectric function, namely, add a periodic
 weak potential coupled with the density operator above and
 compute the ratio of the external versus the effective potential, gives
 the exact RPA dielctric function thereby suggesting that dropping
 terms quadratic in the displacements(in the definition of the density)
 is not going to be important in
 the same limit in which the RPA itself is exact(for a pedagogial description
 of this procedure of obtaining the dielectric function please consult the text
 by Kadanoff and Baym \cite{Baym}). 
 In order to see how good the present theory is, it is desirable to
 compare these results with the Calogero-Sutherland model or
 more specifically with the spin-spin correlation function of the
 Haldane-Shastry model. This is given by Lesage et. al. \cite{Lesage},
  Here they find that
 the momentum distribution is smooth and does not possess any discontinuity
 or kinks. This is in contrast with the solution above (Eq.(~\ref{SOLN})),
  where we see the
 presence of a fermi surface for sufficiently weak repuslsion.
 However, the supersymmetric t-J model\cite{Kura}
 with inverse square interactions does
 posses a jump at the fermi surface, and therefore the present model
 belongs to this latter class.
 Also, the system does not exhibit Wigner
 crystallization at low density. This is another drawback of the interaction
 that we have considered. But this is to be expected since the present model
  captures the two-body repulsive interaction only in the ultra-high
 density limit. On the plus side, the highly nonperturbative
 nature of the solution suggested by the presence of the term, 
 $  cosh( \frac{2 \pi q}{m}\frac{1}{v_{q}} ) - 1  $ cannot be missed.

 It is a pleasure to thank Prof. A. H. Castro-Neto and Prof. D. K. Campbell 
 for providing important references and encouragement and
 Prof. A. J. Leggett for giving his valuable time and advice
 on matters related to the pursuit of this work. Thanks are also due to
 Prof. Ilias E. Perakis for providing the author with an important
 reference and to Prof. Y.C. Chang for general encouragement.
 This work was supported in part by ONR N00014-90-J-1267 and
 the Unversity of Illinois, Materials Research Laboratory under grant
 NSF/DMR-89-20539 and in part by the Dept. of Physics at
 University of Illinois at Urbana-Champaign. The author may be contacted at
 the e-mail address setlur@mrlxpa.mrl.uiuc.edu.

\end{document}